\documentclass[preprint,showpacs,aps,preprintnumbers,amsmath,amssymb]{revtex4-1}

\usepackage{graphicx,epsfig}
\usepackage{dcolumn}
\usepackage{bm}
\newcommand{\be}{\begin{equation}}
\newcommand{\ee}{\end{equation}}
\newcommand{\bse}{\begin{subequations}}
\newcommand{\ese}{\end{subequations}}
\newcommand{\bary}{\begin{eqnarray}}
\newcommand{\eary}{\end{eqnarray}}
\newcommand{\bwt}{\begin{widetext}}
\newcommand{\ewt}{\end{widetext}}

\begin{document}


\title{The nature of the intrinsic spectra from the VHE emission of \\
H 2356-309 and 1ES 1101-232}
\author{Sarira Sahu$^{a,b}$ }
\email{sarira@nucleares.unam.mx}
\author{Alberto Rosales de Le\'{o}n$^{a}$}
\email{albertoros4@ciencias.unam.mx}
\author{Shigehiro Nagataki$^{b,c,d}$}
\email{shigehiro.nagataki@riken.jp}

\affiliation{$^{a}$Instituto de Ciencias Nucleares, Universidad Nacional Aut\'onoma de M\'exico, 
Circuito Exterior, C.U., A. Postal 70-543, 04510 Mexico DF, Mexico}


\affiliation{$^{b}$Astrophysical Big Bang Laboratory, RIKEN, Hirosawa, Wako, Saitama 351-0198, Japan}
\affiliation{$^{c}$Interdisciplinary Theoretical Science Research
  (iTHES), RIKEN, Hirosawa, Wako, Saitama 351-0198, Japan}

\affiliation{$^{d}$Interdisciplinary Theoretical \& Mathematical Science (iTHEMS),
  RIKEN, Hirosawa, Wako, Saitama 351-0198, Japan}


\begin{abstract}

The VHE emission from the HBLs H 2356-309 and 1ES 1101-232 were
observed by HESS telescopes during 2004--2007.
Particularly the observation in 2004 from H 2356-309 and during
2004--2005 from 1ES 1101-232 were analyzed to derive strong upper limits on the EBL
which was found to be consistent with the lower limits from the integrated
light of resolved galaxies. 
Here we have used the photohadronic model corroborated by two
template EBL models to fit the observed VHE gamma-ray data from these
two HBLs and to predict their intrinsic spectra. We obtain very good fit to
the VHE spectra of these two HBLs. However,
the predicted intrinsic spectra are different for each EBL
model. For the HBL H 2356-309, we obtain a flat intrinsic spectrum and
for 1ES 1101-232 the spectrum is mildly harder than 2 but much 
softer than 1.5. 

\end{abstract}

\maketitle

\section{Introduction}
\label{sec:intro}
The high energy $\gamma$-rays coming from the distant blazar jets to the
Earth are attenuated by pair production with the soft
photons\cite{Stecker:1992wi,Hauser:2001xs}. There
are mainly two important sources of these soft photons, namely, 
synchrotron photons intrinsic to the jet and the external ambient photons from the 
extragalactic background light
(EBL). As we understand, the blazar spectra are highly variable and
have wider range of variability. Although we have learned a lot about them, the
present understanding of their radiation process is still incomplete 
to reliably predict the intrinsic TeV spectrum, and thus to disentangle absorption from intrinsic
features. It is hoped that  modeling of the blazar spectral energy distribution (SED) by taking into account
properly  the emission mechanism can take care of the intrinsic
extraneous effect due to its environment.
The total absorption of the TeV $\gamma$-rays depends on the local
density of the low energy photons at the origin, the distance traveled (redshift
$z$) and also the energy of the high energy $\gamma$-rays $E_{\gamma}$. For higher
energy $\gamma$-rays the absorption process leads to the steepening of
the observed spectrum thus reducing the observed flux.
So the observed blazar spectrum contains valuable information about the
history of EBL in the line-of-sight and the intrinsic properties of the source.

The EBL effect on the blazar spectrum can be
calculated by subtracting the foreground sources from the diffuse
emission. However, the foreground zodiacal light and galactic light
introduce large uncertainties in such measurements and make it
difficult to isolate the EBL contribution from the observed multi-TeV
flux from distant blazars. Strict lower limits are derived from the
source counts and rather loose upper limits come from direct
measurements. Nevertheless, an indirect approach is to utilize the very high energy (VHE) $\gamma$-ray
spectra from blazars by assuming a power-law behavior for the intrinsic spectrum.
So, long term studies of many high frequency peaked BL Lacerate objects
(HBLs) of different redshifts during periods of activity such as
flaring will provide invaluable insights into the emission
mechanisms responsible for the production of VHE $\gamma$-rays
as well as the absorption process due to EBL. In recent years, the
continuing success of highly sensitive Imaging Atmospheric Cherenkov
Telescopes (IACTs) such as VERITAS\cite{Archambault:2016wly}, HESS \cite{Abramowski:2012ry} and MAGIC\cite{Moralejo:2017eny} have led
to the discovery of many new extragalactic TeV sources which in turn
resulted in constraining  the flux density of the EBL over two decades
of wavelengths from $\sim 0.30\, \mu$m to
$17\,\mu$m\cite{H.E.S.S.:2017odt,Romoli:2017nrp,Moralejo:2017eny, Abeysekara:2015pjl,Ahnen:2015qrv,Ahnen:2016gog}. 

Blazars detected at VHE are predominantly HBLs and flaring in VHE seems to be a
common phenomenon in these objects, although it is not
yet understood properly. In general this VHE emission is explained by
leptonic models\cite{Katarzynski:2006db,Fossati:1998zn,Ghisellini:1998it,Roustazadeh:2011zz}
through SSC scattering process. 
Due to the absorption of the primary
VHE photons by EBL, the corresponding intrinsic spectrum becomes
harder than the observed one. Normally in the SSC model the
intrinsic photon spectrum has a spectral index
$\alpha_{int} > 1.5$ (discussed in Sec.\ref{sec:flaremodel}) in the energy range where electron cooling via synchrotron
and/or IC energy loss is efficient and the hard spectrum with $\alpha_{int}= 1.5$
is considered as a lower bound. It is difficult to  produce
harder spectra ($\alpha_{int} < 1.5$)  in the one-zone SSC scenario. 
The orphan flaring in multi-TeV $\gamma$-rays and  blazars with
hard gamma ray spectra are troublesome to
deal with the standard SSC scenario. Multi-TeV emission from 
two  HBLs, 1ES 1101-232 (z=0.186) and H 2356-309 (z=0.165) were
observed by the HESS Cherenkov telescopes\cite{Aharonian:2005gh}
and at that time these were the most distant sources.
Due to the lack of reliable EBL data, different EBL SEDs were assumed
to construct the intrinsic spectra from the observed VHE spectra.
The assumed EBL SEDs were in general agreement with the EBL
spectrum expected from galaxy emission. Although,  
the constructed intrinsic spectra were compatible with a
power-law, the intrinsic spectrum of the HBL 1ES
1101-232 was rather hard and
such hard spectra had never been
observed before in the spectra of closest, less absorbed TeV blazars e.g. Mrk 421
and Mrk 501\cite{Aharonian:1999vy,Krennrich:2002as,DjannatiAtai:1999af,Sahu:2015tua,Sahu:2016bdu} and are difficult to explain with the standard leptonic
or hadronic scenarios\cite{Aharonian:2001jg} for blazar emission. Also
the resulting EBL upper limits were found to be consistent with the
lower limits from the integrated light of resolved galaxies and seems
to exclude a large contribution to the EBL from other sources. From
the analysis in ref. \cite{Aharonian:2005gh}, it was inferred that the Universe is more transparent to
gamma rays than previously anticipated. Later on, harder spectra
have also been observed from many HBLs \cite{Aharonian:2007wc,Neronov:2011ni,HESS:2012aa}.
Thereafter, many scenarios are suggested to achieve very hard VHE
spectra which are discussed in ref. \cite{Cerruti:2014iwa} and
references therein. Also alternative photohadronic scenarios are
proposed to explain the VHE emission\cite{Essey:2010er,Cao:2014nia}. The structured jet
(spine-layer) model is also proposed to explain the  high energy
emission from blazars\cite{Ghisellini:2004ec,Tavecchio:2008be}.

In this work our goal is to use the photohadronic model of Sahu et al.\cite{Sahu:2013ixa} 
and different template EBL models\cite{Franceschini:2008tp,Inoue:2012bk} to re-examine
the VHE spectra of HBLs 1ES 1101-232 and H 2356-309 and to calculate
their intrinsic spectra. Here, we assume that
the Fermi accelerated protons in the blazar jet have a power-law
behavior and the observed VHE spectra of the HBLs
are related to the proton spectrum. 

The paper is organized as follows: In Sec. 2 we discuss 
different EBL models which are used for our calculation. The
photohadronic model of Sahu et al.\cite{Sahu:2013ixa} is discussed concisely in Sec. 3.
We discuss the results obtained for the VHE observations of HBLs  H
2356-309 and 1ES 1101-232 in Sec.4 and finally we briefly summarize our results in Sec. 5.

\section{EBL Models}
\label{sec:eblmodels}

Considering the uncertainty associated with the direct detection of
the EBL contribution, a wide range of models have been developed to model
the EBL SED based on our knowledge of galaxy and star formation rate
and at the same time incorporating the observational
inputs\cite{Franceschini:2008tp,Inoue:2012bk,Kneiske:2002wi,Stecker:2005qs,Dominguez:2010bv,Orr:2011ha,Primack:2005rf}. Mainly
three types of EBL models exist: backward and forward evolution models
and semi-analytical galaxy formation models with a combination of
information about galaxy evolution and
observed properties of galaxy spectra.
In the backward evolution scenarios\cite{Stecker:2005qs},  one starts from the observed
properties of galaxies in the local universe and evolve them from cosmological initial conditions or
extrapolating backward in time using parametric models of the
evolution of galaxies. This extrapolation induces
uncertainties  in the properties of the EBL which increases at high redshifts.
However, the forward evolution models\cite{Franceschini:2008tp,Kneiske:2002wi} predict the temporal evolution of
galaxies forward in time starting from the cosmological
initial conditions. Although, these models are successful in
reproducing the general characteristics of the observed EBL, cannot
account for the detailed evolution of  important quantities such as
the metallicity and dust content, which can significantly affect the
shape of the EBL. Finally, semi-analytical models have been developed
which follow the formation of large scale structures driven by cold dark
matter in the universe by using the cosmological parameters from
observations. This method also accounts for the merging of the dark matter
halos and the emergence of galaxies which form as baryonic matter falls into the
potential wells of these halos. Such models are successful in
reproducing observed properties of galaxies from local universe up to
$z\sim 6$.

\begin{figure}
{\centering
\resizebox*{0.5\textwidth}{0.35\textheight}
{\includegraphics{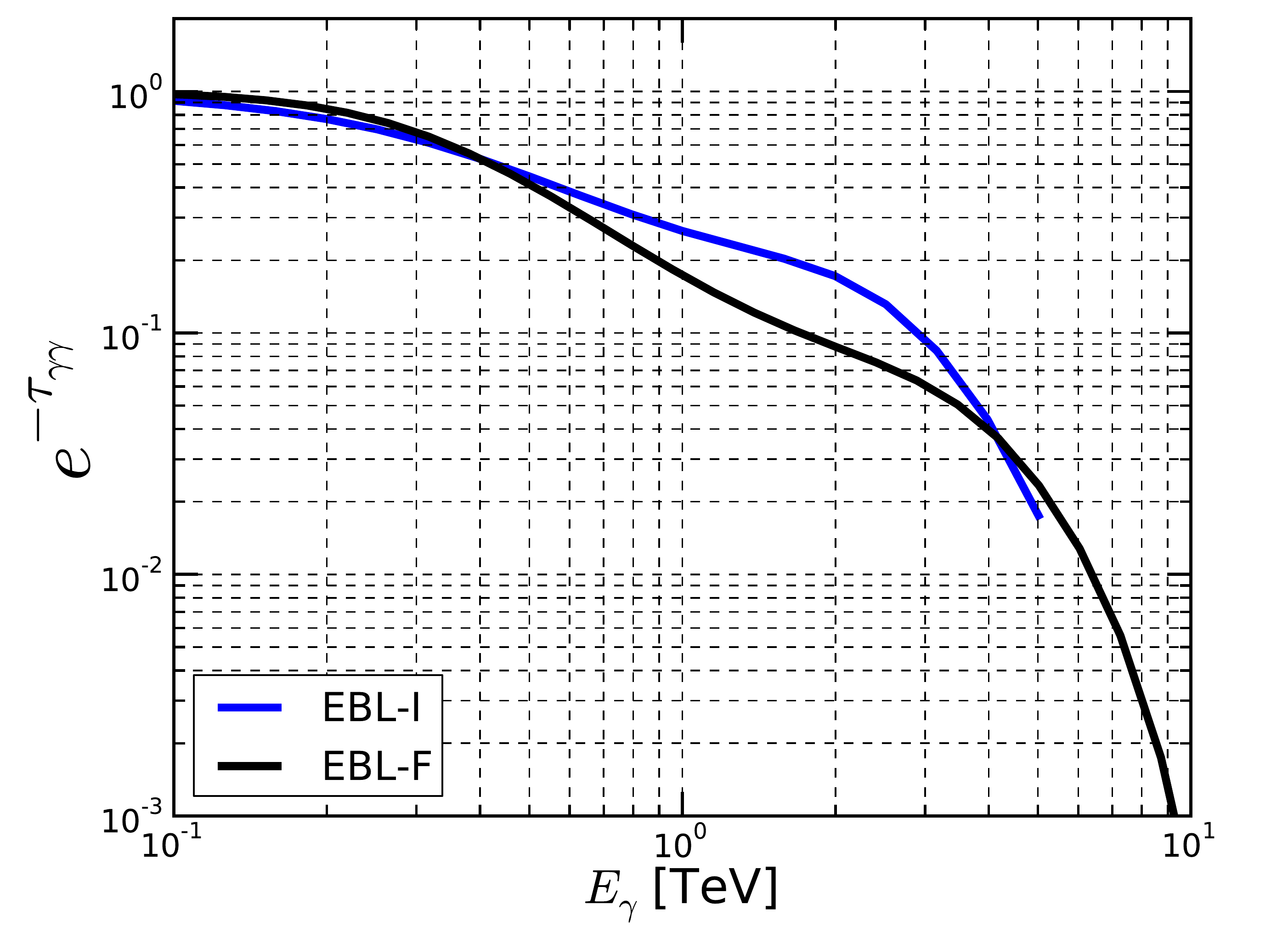}}
\par}
\caption{
At a redshift of $z=0.165$, the attenuation factor
$e^{-\tau_{\gamma\gamma}}$ as a function of VHE $\gamma$-ray energy 
$E_{\gamma}$ for different EBL models are shown for comparison. 
\label{fig:EBL309}
}
\end{figure}
\begin{figure}
{\centering
\resizebox*{0.5\textwidth}{0.35\textheight}
{\includegraphics{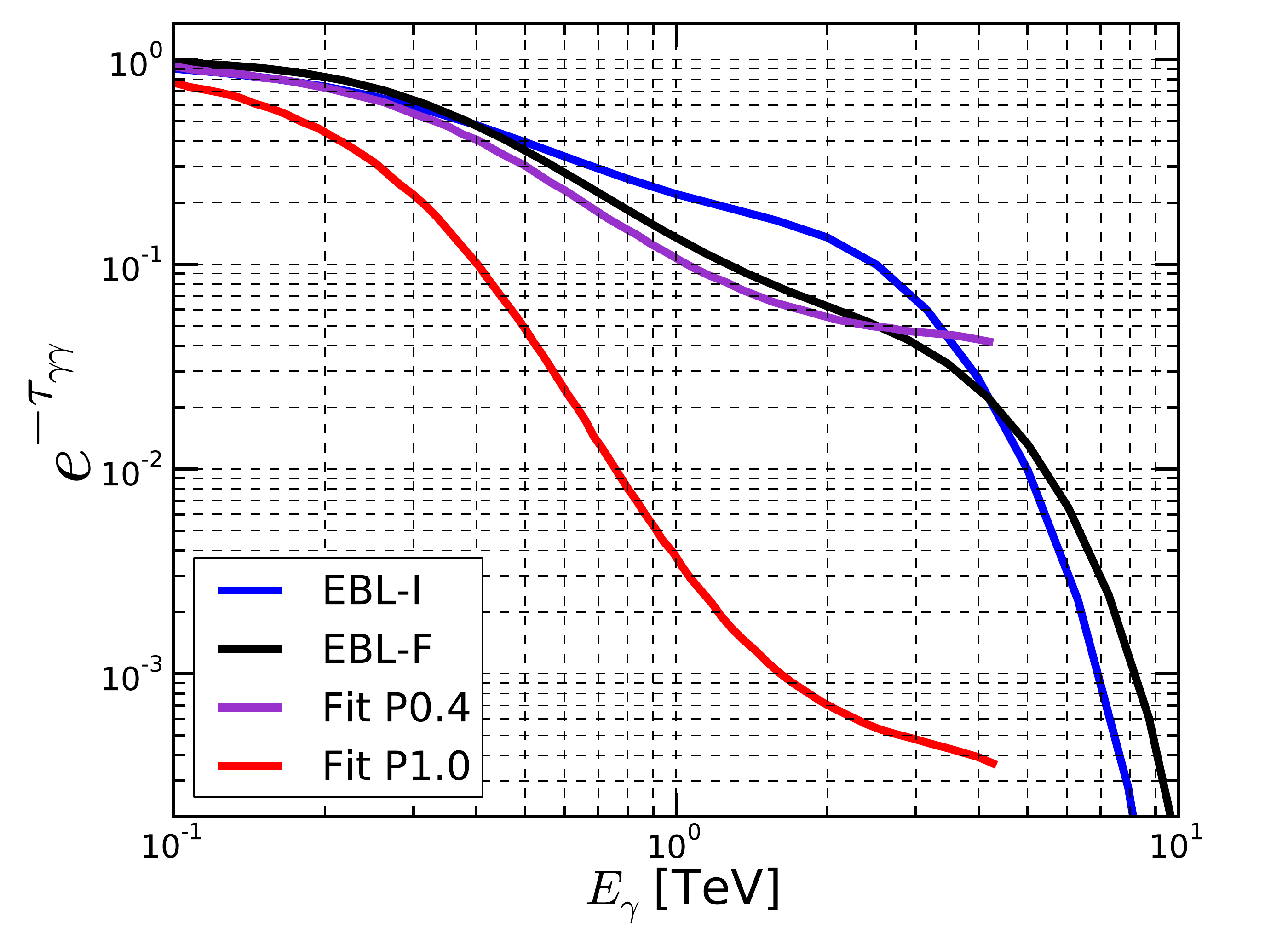}}
\par}
\caption{
At a redshift of $z=0.186$, the attenuation factor
$e^{-\tau_{\gamma\gamma}}$ as a function of VHE $\gamma$-ray energy 
$E_{\gamma}$ for different EBL models are shown for comparison. 
The attenuation factors of ref. \cite{Aharonian:2005gh} are labeled
as P1.0 and P0.4 which correspond to originally normalized to match
the direct estimate (P1.0) and scaled by a factor 0.40 (P0.4) respectively. 
\label{fig:EBL232}
}
\end{figure}


The VHE $\gamma$-rays from distant sources interact with the
EBL to produce electron-positron
pairs thus depleting the VHE flux by a factor of
$e^{-\tau_{\gamma\gamma}}$. Here $\tau_{\gamma\gamma}$ is the 
optical depth of the process
$\gamma\gamma\rightarrow e^+e^-$ which depends on the  energy
of the $\gamma$-ray ($E_{\gamma}$)  and the redshift (z).
For the present study we choose two different EBL models
by Franceschini et al.\cite{Franceschini:2008tp} and Inoue et
al.\cite{Inoue:2012bk} (hereafter EBL-F and EBL-I respectively). The
attenuation factor $e^{-\tau_{\gamma\gamma}}$ of these EBL models at redshifts 0.165
and 0.186 are shown in Figs.\ref{fig:EBL309} and \ref{fig:EBL232}. 
In Fig.\ref{fig:EBL309}, it is observed that below $\sim 500$ GeV,
these two models behave almost the same and above this energy there is
a slight difference in their behavior.


In Fig.\ref{fig:EBL232}, we have compared the attenuation factor of
EBL-F and EBL-I for z=0.186. The behavior of both these models are
similar to the one at redshift z=0.165 which can be seen by comparing 
Fig.\ref{fig:EBL309} and \ref{fig:EBL232}. Here in Fig.\ref{fig:EBL232} we have also plotted the
attenuation factor of ref. \cite{Aharonian:2005gh} for z=0.186 with two different
normalizations. 
To determine an upper limit of the EBL model, Aharonian et al [16]
assumed a previously known shape for the SED of the EBL . This curve,
is then renormalized to fit the measurements made by the HESS
collaboration at 2.2 and 3.5 $\mu$m. Here the normalization factor is
as a free parameter and the
scaled curves are named accordingly to this factor. Here,  P1.0 means the
original shape is multiply by a factor of 1, meanwhile P0.4 means that
the original shape of the EBL is scaled by a factor of 0.4. In Fig.\ref{fig:EBL232} the P0.4
curve (violet) is very similar to the EBL-F
(black curve) up to $E_{\gamma}\sim 2$ TeV. The curve P1.0 (red)
falls very fast compared to other, as can be seen in the figure. 
This fast fall corresponds to a denser EBL component.

\section{Photohadronic Model}
\label{sec:flaremodel}
The photohadronic model \cite{Sahu:2013ixa,Sahu:2015tua,Sahu:2016bdu,Sahu:2013cja} 
has explained  very well the orphan TeV flare
from the blazar 1ES1959+650, and multi-TeV emission from M87, Mrk 421, Mrk 501 and
1ES 1011+496. This  model relies on
the standard interpretation of the leptonic model to explain both low and
high energy peaks by synchrotron and SSC photons respectively, as in the
case of any other AGN and blazar. Thereafter, it is assumed that the
flaring occurs within a compact and confined volume of size $R'_f$
inside the blob of radius $R'_b$  (where $^{\prime}$ implies the jet
comoving frame) and $R'_f <R'_b$ .  During the flaring, 
both the internal and the external jets are moving with almost the
same bulk Lorentz factor $\Gamma_{in}\simeq \Gamma_{ext}\simeq \Gamma$
and the Doppler factor ${\cal  D}$ as the blob (for blazars
$\Gamma\simeq {\cal  D}$).  A detail description of the
photohadronic model and its geometrical structure
is discussed in ref.\cite{Sahu:2013ixa}.
The injected spectrum of the
Fermi accelerated charged particles having a power-law spectrum 
$dN/dE\propto E^{-\alpha}$ with the power index $\alpha \ge 2$ is
considered here. 

In the compact inner jet region, the Fermi accelerated high energy protons 
interact with the background photons with a
comoving density $n'_{\gamma,f}$ to produce the $\Delta$-resonance
and its subsequent decay to neutral and charged pions will give
VHE $\gamma$-rays and neutrinos respectively.
In the flaring region we assume
$n'_{\gamma,f}$ is much higher than
the rest of the blob $n'_{\gamma}$ (non-flaring) i.e.
${n'_{\gamma, f}(\epsilon_\gamma)}\gg
{n'_{\gamma}(\epsilon_\gamma)}$. As the inner jet is buried within the
blob, we can't calculate $n'_{\gamma,f}$ directly. So we use the
scaling behavior of the photon densities in 
the inner and the outer jet regions as follows:
\be
\frac{n'_{\gamma, f}(\epsilon_{\gamma_1})}
{n'_{\gamma, f}(\epsilon_{\gamma_2})} \simeq \frac{n'_\gamma(\epsilon_{\gamma_1})}
{n'_\gamma(\epsilon_{\gamma_2})},
\label{densityratio}
\ee 
which assumes that
the ratio of photon densities at two different
background energies $\epsilon_{\gamma_1} $  and $\epsilon_{\gamma_2} $
in the flaring and the non-flaring states remains almost the same. The
photon density in the outer region can be calculated from the observed
flux from SED. By using Eq. (\ref{densityratio}),  the
$n'_{\gamma,f}$ can be expressed in terms of $n'_{\gamma}$. 
It is shown in Refs. \cite{Cao:2014nia,Zdziarski:2015rsa} that super Eddington
luminosity in protons is required to explain the high energy
peaks. In a normal jet, the photon density is low, which makes the
photohadronic process inefficient\cite{Pjanka:2016ylv}. However, in the present
scenario it is assumed that during the flaring the photon density in
the inner jet region can go up so that the $\Delta$-resonance
production is moderately efficient, which eliminates the extreme
energy requirement\cite{Sahu:2015tua}.

In the observer frame, the $\pi^0$-decay TeV photon energy
$E_{\gamma}$ and the 
target photon energy $\epsilon_{\gamma}$ 
satisfy the condition
\be
E_\gamma \epsilon_\gamma \simeq 0.032\,\, \frac{{\cal D}^2}{ (1+z)^{2}} ~{\rm GeV}^2.
\label{Eegamma}
\ee
The above condition is derived from the process $p\gamma\rightarrow\Delta$.
Also, the observed TeV $\gamma$-ray energy and the proton energy $E_p$ are related through
$E_p\simeq 10\,E_{\gamma}$.
It is observed that for most of the  HBLs, the ${\cal D}$ is such that,
$\epsilon_{\gamma}$ always lies in the lower tail region of the
SSC band. So it is the low energy SSC region which is responsible for the
production of multi-TeV $\gamma$-rays in the photohadronic model.
The efficiency of the  $p\gamma$ process depends on the
physical conditions of the interaction region, such as the size, 
the distance from the base of the jet, expansion time scale (or
dynamical time scale of the blob $t'_d=R'_f$) and the photon density
in the region which is related to the optical depth $\tau_{p\gamma}$ of this process.

Correcting for the EBL contribution,
the observed VHE flux $F_{\gamma}$ can be
expressed in terms of the intrinsic flux $F_{\gamma,int}$ by the relation
\be
F_{\gamma}(E_{\gamma}) = F_{\gamma,int}(E_{\gamma})
e^{-\tau_{\gamma\gamma}(E_{\gamma},z)},
\label{fluxrelat}
\ee
where the intrinsic flux can be given as\cite{Sahu:2016bdu}
\be
F_{\gamma,int}(E_{\gamma})=A_{\gamma} \Phi_{SSC}(\epsilon_{\gamma} )
  \left (\frac{E_{\gamma}}{TeV}\right )^{-\alpha+3}.
\label{fint}
\ee
The SSC energy $\epsilon_{\gamma}$ and the observed energy
$E_{\gamma}$ satisfy the kinematical condition given in
Eq.(\ref{Eegamma})
and $\Phi_{SSC}$ is the SSC flux corresponding to
the energy $\epsilon_{\gamma}$ which is known from the leptonic model
fit to the multi-wavelength data.
Here the only free parameter is the spectral
index $\alpha$. For a given multi-TeV flaring energy and its
corresponding flux, we can always
look for the best fit to the spectrum which will give the value of
$A_{\gamma}$. Also it is to be noted that, blazars are highly variable
objects and characterized by very wide range of different spectra. Our
model depends on the value of $\Phi_{SSC}$ which can be different for
separate epochs of observations and accordingly the value of
$A_{\gamma}$ can vary. However, in principle $\alpha$ should be kept
constant for a given acceleration mechanism.
In the leptonic model, the SSC photon flux in the low energy tail
region is a power-law given as $\Phi_{SSC} \propto
\epsilon^{\beta}_{\gamma}$, where $\beta > 0$. By using the relation
in Eq. (\ref{Eegamma}) we can express $\epsilon_{\gamma}$ in terms of the observed
VHE $\gamma$-ray energy $E_{\gamma}$ which will give $\Phi_{SSC} \propto
E^{-\beta}_{\gamma}$ and again by replacing $\Phi_{SSC}$ in
Eq.(\ref{fint}) we get
\be
F_{\gamma,int}(E_{\gamma})\propto 
 \left (\frac{E_{\gamma}}{TeV}\right )^{-\alpha-\beta+3},
\ee
and the intrinsic differential power spectrum for VHE photon is a
power-law given as
\be
\left (\frac{dN}{dE_{\gamma} } \right )_{int} \propto  
 \left (\frac{E_{\gamma}}{TeV}\right )^{-\alpha_{int}}
\,\, \, {\text {with}}\, \, \, \alpha_{int}={\alpha+\beta-1}.
\ee
However, due to the nonlinearity of 
${\tau_{\gamma\gamma}}$ the observed VHE flux will not behave as a single
power-law.
Hardness of the intrinsic spectrum depends on the value of $\alpha$
for a given leptonic model which fixes the value of $\beta$.

\begin{table}
\centering
\caption{The following parameters are taken from the one-zone
  leptonic model of ref. \cite{Abramowski:2010kg} to fit the SED of
H 2356-309  and from ref. \cite{Costamante:2006pm}
to fit the SED of 1ES 1101-232. 
} 
\begin{tabular*}{\columnwidth}{@{\extracolsep{\fill}}llll@{}}
\hline
\multicolumn{1}{@{}l}{Parameter} &Description & H 2356-309 & 1ES 1101-232\\
\hline
\hline
$M_{BH}$ & Black hole mass & $\sim 10^9 M_{\odot}$ & $\sim 10^9 M_{\odot}$\\
z & Redshift & 0.165  & 0.186\\
$\Gamma$ &Bulk Lorentz Factor & 18 & 20\\
${\cal D}$& Doppler Factor & 18 & 20\\
$R^{\prime}_b$ & Blob Radius & $7.5\times 10^{15}\, cm$ & $10^{16}\,cm$\\ 
$B^{\prime}$ &Magnetic Field & 0.16\, G & 0.1 \, G\\ 
\hline
\end{tabular*}
\label{tab:tab1}
\end{table}
\begin{figure}
{\centering
\resizebox*{0.5\textwidth}{0.35\textheight}
{\includegraphics{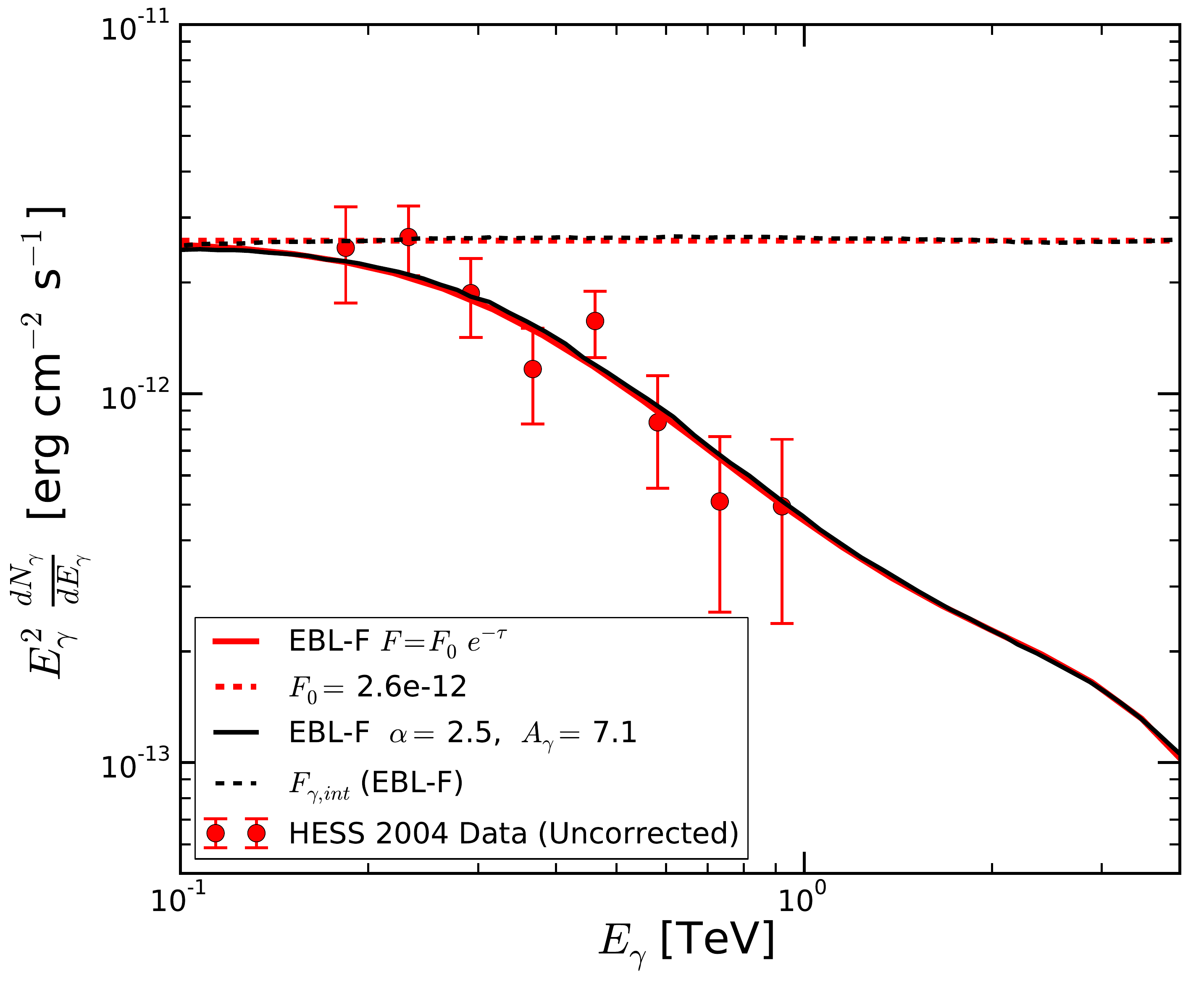}}
\par}
\caption{The VHE emission from HBL H 2356-309 during 2004 observation
  by HESS telescopes is shown along with the rescaling of the
  EBL-F attenuation factor by a constant $F_0=2.6\times 10^{-12}\,
  erg\, cm^{-2}\, s^{-1}$ and EBL-F correction to the photohadronic
  model ($\alpha=2.5$ and $A_{\gamma}=7.1$). The intrinsic fluxes are also shown. 
\label{fig:309eblFran}
}
\end{figure}

\begin{figure}
{\centering
\resizebox*{0.5\textwidth}{0.35\textheight}
{\includegraphics{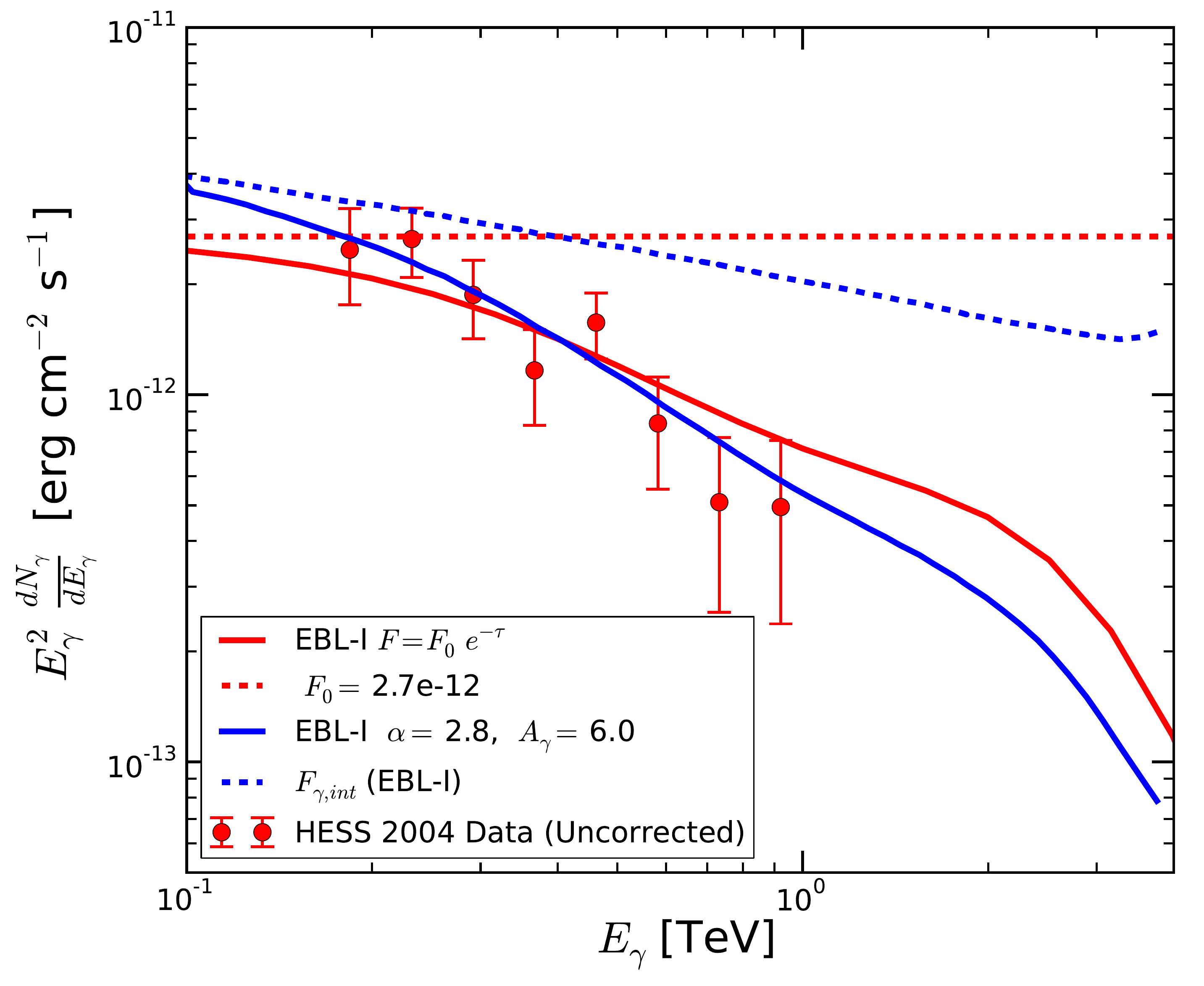}}
\par}
\caption{The VHE emission from HBL H 2356-309 during 2004 observation
  by HESS telescopes is shown along with the rescaling of the
  EBL-I attenuation factor by a constant $F_0=2.7\times 10^{-12}\,
  erg\, cm^{-2}\, s^{-1}$ and EBL-I correction to the photohadronic
  model ($\alpha=2.8$ and $A_{\gamma}=6.0$). The intrinsic fluxes are also shown.
\label{fig:309eblInoue}
}
\end{figure}

\begin{figure}
{\centering
\resizebox*{0.5\textwidth}{0.35\textheight}
{\includegraphics{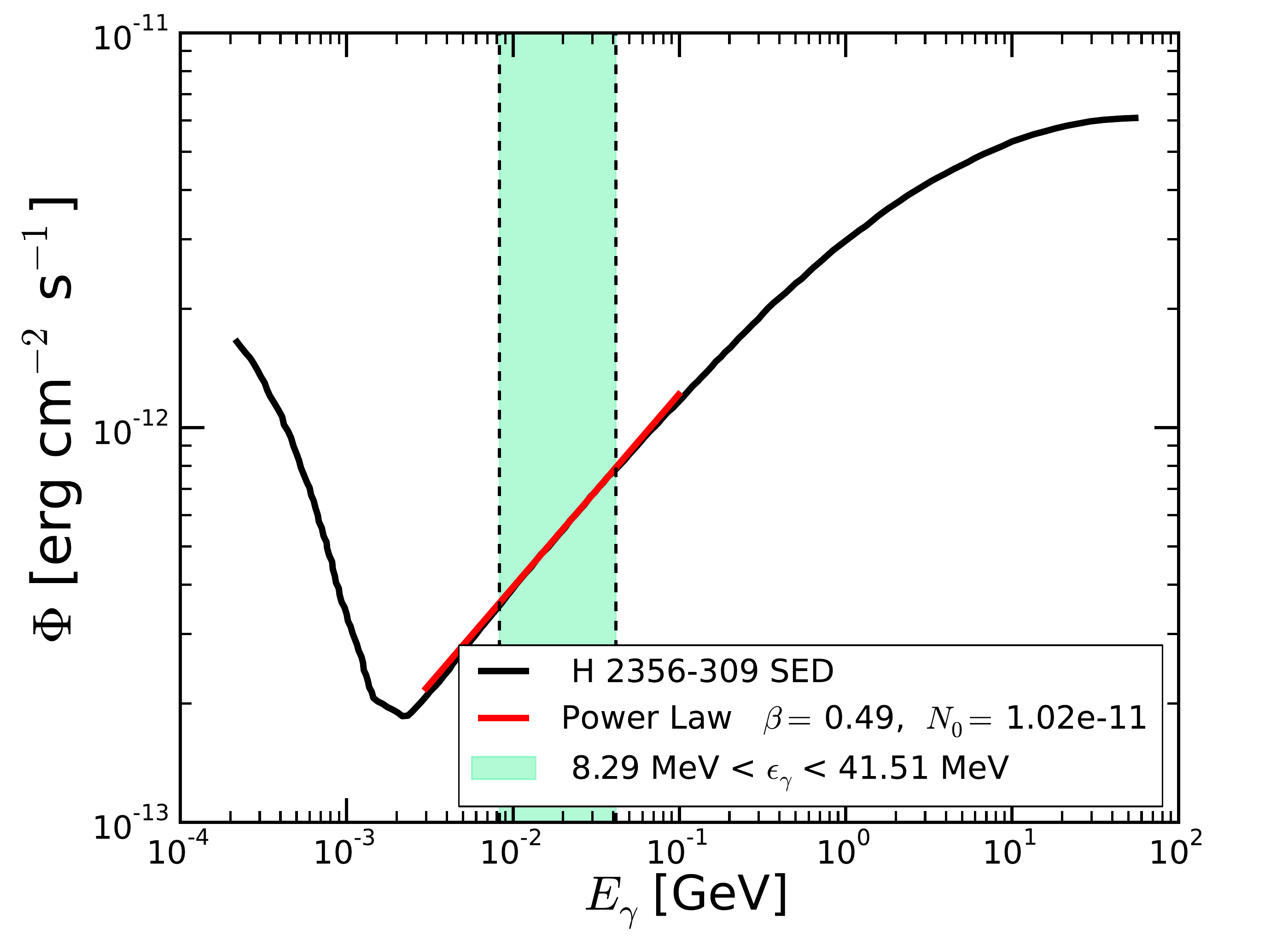}}
\par}
\caption{ 
The leptonic SED of the HBL H 2356-309 fitted with one-zone leptonic
model is from ref. \cite{Abramowski:2010kg} and the
shaded region in the figure is the 
SSC energy range $8.3\, MeV (2\times 10^{21}\, Hz\le \epsilon_{\gamma}
\le 41.5\, MeV (1\times 10^{22}\, Hz)$ corresponding to the VHE
$\gamma$-ray energy in the range $0.18\, TeV \le E_{\gamma}\le 0.92\,
TeV$ of the HBL H 2356-309. The corresponding SSC flux $\Phi_{SSC}$ in y-axis
is fitted with a power-law 
$\Phi_{SSC}=N_0 E^{-\beta}_{\gamma,{\text {TeV}}}$, where $N_0=1.02\times 10^{-11}\,
erg\, cm^{-2}\, s^{-1}$, $\beta=0.49$ and $E_{\gamma,{\text {TeV}}}$ implies $E_{\gamma}$
is expressed in units of TeV.  
\label{fig:309sscpowerlaw}
}
\end{figure}

\begin{figure}
{\centering
\resizebox*{0.5\textwidth}{0.35\textheight}
{\includegraphics{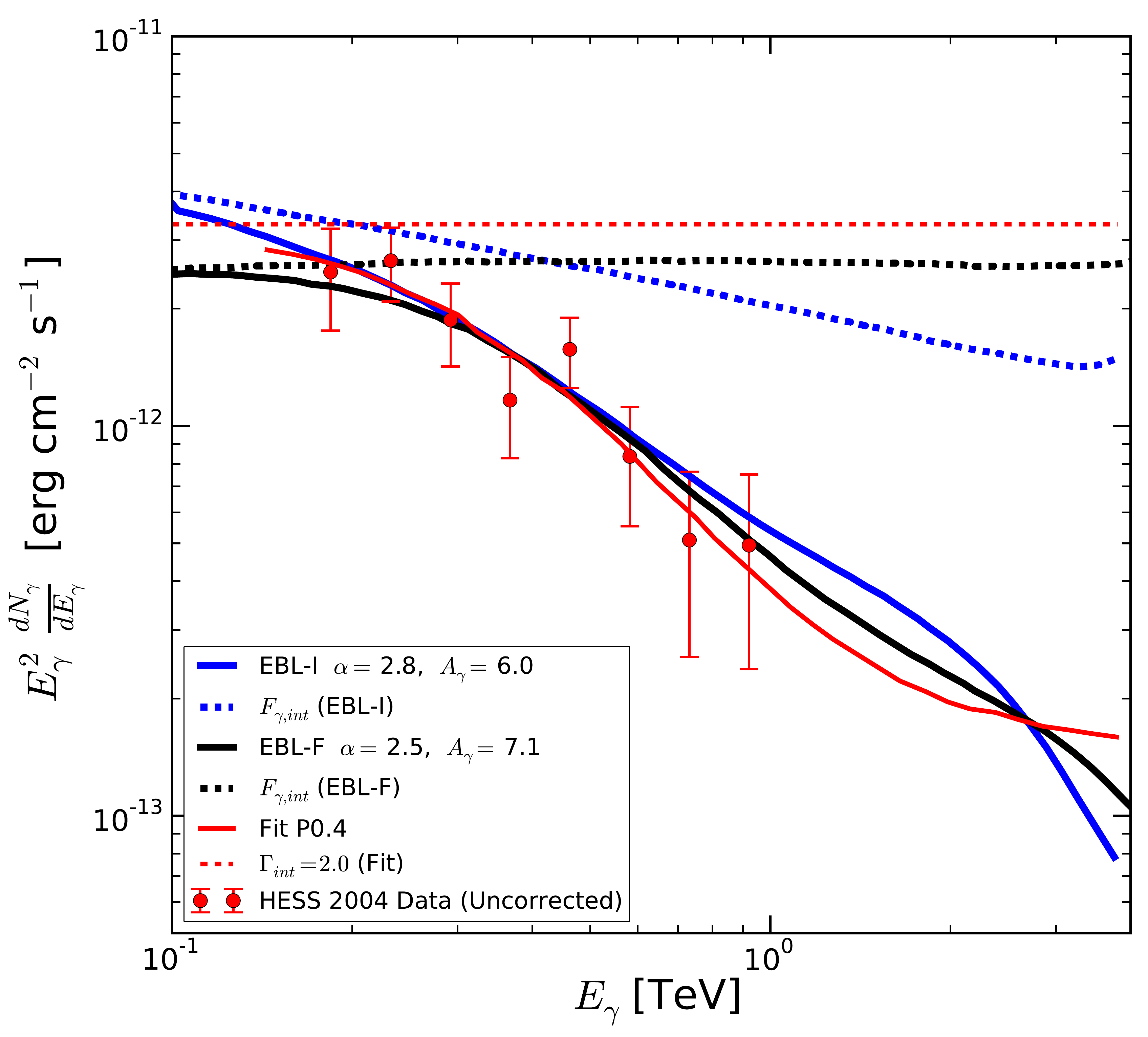}}
\par}
\caption{The observed VHE flux from H 2356-309 is fitted using EBL-F and EBL-I to the
  photohadronic model. The fit to the data in ref. \cite{Aharonian:2005gh}
 using
  P0.4 is shown for comparison. The intrinsic flux predicted by 
  these models are also shown, where in the above reference
  the intrinsic flux is fitted with
  a power-law give by $F_{\gamma,int}=3.3\times
  10^{-12}\,(E_{\gamma}/TeV)^{-\Gamma_{int}}\,erg\,cm^{-2}\, s^{-1}$
  with $\Gamma_{int}=2.0$.
\label{fig:SED309}
}
\end{figure}

\section{Results}

The VHE emission from the HBLs H 2356-309 and 1ES 1101-232 were 
observed by HESS telescopes during 2004 and 2005. The intrinsic flux
can be calculated from the observed one by subtracting the EBL
effect. So we use reliable EBL models to calculate the intrinsic
flux. For our interpretation of the VHE $\gamma$-ray spectrum, we use
two EBL models: EBL-F and EBL-I which are discussed in Sec.\ref{sec:eblmodels}.
Also, we have to model the emission process in the HBLs. So
here we use the photohadronic model of Sahu et al.\cite{Sahu:2013ixa} and input for the
photohadronic process comes from the leptonic models which are
successful in explaining the double peak structure of the blazars. The detail
analysis and results of the HBLs H 2356-309 and 1ES 1101-232 are
discussed separately below.

\subsection{H 2356-309}
The high frequency peaked BL Lac object H 2356-309
is hosted by an elliptical
galaxy located at a redshift of z = 0.165 \cite{Falomo:1991a} and was first
detected in X-rays by the satellite experiment
UHURU\cite{Forman:1978a} and subsequently 
by the Large Area Sky Survey experiment
onboard the HEAO-I satellite \cite{Wood:1984a}. Also in optical band
it was observed\cite{Schwartz:1989}. In 2004, H 2356-309 was
observed simultaneously in X-rays by RXTE, in optical by ROTSE-III, in
radio by Nancay decimetric telescope (NRT) and in VHE for about 40
hours (June to December 2004) by HESS telescopes. It was observed that during this period, the
X-ray spectrum measured above 2 eV was softer and the flux was $\sim
3$ times lower than the one measured by BeppoSAX in 1998 in the same
energy band but in a comparatively quiescent state. Since 2004,
H 2356-309 has been monitored by HESS for several years (from 2005 to
2007) and little flux variability is observed on the time scale of a
few years. From the above simultaneous multi-wavelength observations, the HESS
collaboration used one-zone leptonic model to fit the observed data\cite{Abramowski:2010kg,Aharonian:2006zy} and the best fit parameters of the model are given in
Table-\ref{tab:tab1} which are used for the photohadronic model.

During 2004, the
HESS telescopes observed the VHE emission in the energy range
$0.18\, TeV \le E_{\gamma}\le 0.92\, TeV$ \cite{Aharonian:2005gh} which was
analyzed to constraint the EBL contribution. Here we would like to
mention that the photohadronic model is applicable not only to VHE
flaring but also to VHE (multi-TeV) emission from the blazars under
discussion. In the photohadronic scenario
this range of $E_{\gamma}$ corresponds to Fermi accelerated protons in the energy range
$1.8\, TeV \le E_p \le 10$ TeV which interacts with the seed SSC photons in the
inner jet region in the energy range $8.3\, MeV (2\times 10^{21}\, Hz\le \epsilon_{\gamma}
\le 41.5\, MeV (1\times 10^{22}\, Hz)$ to produce
$\Delta$-resonance. Subsequent decay of the resonance state produces
$\gamma$-rays and neutrinos. 

The VHE spectrum of H 2356-309 is strongly affected  by the EBL and to
calculate the intrinsic spectrum, we have used the EBL-F and EBL-I. 
Observed flux is proportional to the attenuation factor as shown in
Eq.(\ref{fluxrelat}) and by assuming $F_{\gamma,int}$
a constant in both the EBL models, we tried
to fit the observed data which are shown in Figs. \ref{fig:309eblFran}
and \ref{fig:309eblInoue} respectively. It is observed that by  taking
$F_{\gamma,int}=2.6\times 10^{-12}\, erg\, cm^{-2}\,s^{-1}$ for EBL-F
(red curve) we can
fit the observed data very well which is shown in
Fig. \ref{fig:309eblFran}. In the same plot we have also shown the
photohadronic fit (black curve). The photohadronic fit and the multiplication by a
constant factor are indistinguishable. A good fit to the data is
obtained in photohadronic model for  
$\alpha=2.5$ and $A_{\gamma}=7.1$. The constant $F_{\gamma,int}$ implies that
$\beta\simeq 0.5$ and exact fit to the $\Phi_{SSC}$  in the energy
range $8.3\, MeV \,\le \epsilon_{\gamma}\le 41.5\, MeV$ gives
$\beta=0.49$ which is shown in Fig. \ref{fig:309sscpowerlaw} (red line). Also
this gives the spectral index $\alpha_{int}\simeq 2$ for the intrinsic
spectrum\cite{Costamante:2012ph}.

In Fig. \ref{fig:309eblInoue} we have also rescaled the attenuation
factor of EBL-I (red curve) by 
$F_{\gamma,int}=2.7\times 10^{-12}\, erg\, cm^{-2}\,s^{-1}$ 
to fit the observed VHE data and for comparison
the photohadronic model fit (blue curve) is also shown.
The best fit for
the photohadronic model is achieved here for $\alpha=2.8$ and
$A_{\gamma}=6.0$. We observe that the rescaling and the model fit are
very different from each other and the photohadronic model fit is
better than the rescale one. We also observe that the EBL-I (blue curve) correction
to the photohadronic fit does not give a constant  
$F_{\gamma,int}$, but a power-law with $F_{\gamma,int}\propto E^{-0.3}_{\gamma}$ and
the intrinsic spectral index is $\alpha_{int}\simeq 2.3$.

To compare the predictions of different EBL models and the result of ref. \cite{Aharonian:2005gh}
with P0.4 scaling (red curve), we have plotted these results in Fig. \ref{fig:SED309}. We
observe that all these models fit well to the observed data. For
$E_{\gamma} < 300$ GeV the EBL-F (black curve) predict slightly lower flux than
the rest. Also for $E_{\gamma} > 2.7$ TeV these predictions slightly differ from
each other. Although the EBL-F (black dotted curve) and ref. \cite{Aharonian:2005gh} models predict flat
$F_{\gamma,int}$, their magnitudes are different due to different
normalizations. The EBL-I  predict an intrinsic flux with soft power
index $F_{\gamma,int}\propto E^{-0.3}_{\gamma}$ (blue dotted curve). 

The Bethe-Heitler (BH) pair production process $p\gamma\rightarrow
pe^+e^-$ can also compete with the photohadronic process, but strongly
depends on the angle between the photon and the emitted leptons. In the BH
process, the electron-position pair can emit synchrotron photons. It
is shown that this process can produce a third peak in-between the
synchrotron peak and the IC peak\cite{Petropoulou:2014rla}.
For this to happen, the protons and electrons energies have to be 
very high\cite{Petropoulou:2015upa}. It the
present scenario, the maximum energy of a proton and also an
electron in the jet is $\sim 10$ TeV. For a magnetic field of 
0.16 G, an electron of energy 10 TeV will emit a synchrotron photon 
with maximum energy $\epsilon_{\gamma}\sim 0.8$ MeV, which is an order of
magnitude smaller than the lowest SSC photon energy $\epsilon_{\gamma}=
8.3$ MeV taking part in the photohadronic process to produce
$\Delta$-resonance. Leptons produced from pion and muon decay, pair creation and
BH process will
have energies less than 10 TeV and again the synchrotron photons from these
leptons will have energies less than 0.8 MeV. 
The BH process may be important for very high
energy protons and electrons, but here it does not play an important role
and will not enhance the SSC photon flux in the energy range $8.3\,
MeV \,\le \epsilon_{\gamma}\le 41.5\, MeV$ unless the magnetic field
is high. 

\begin{figure}
{\centering
\resizebox*{0.5\textwidth}{0.35\textheight}
{\includegraphics{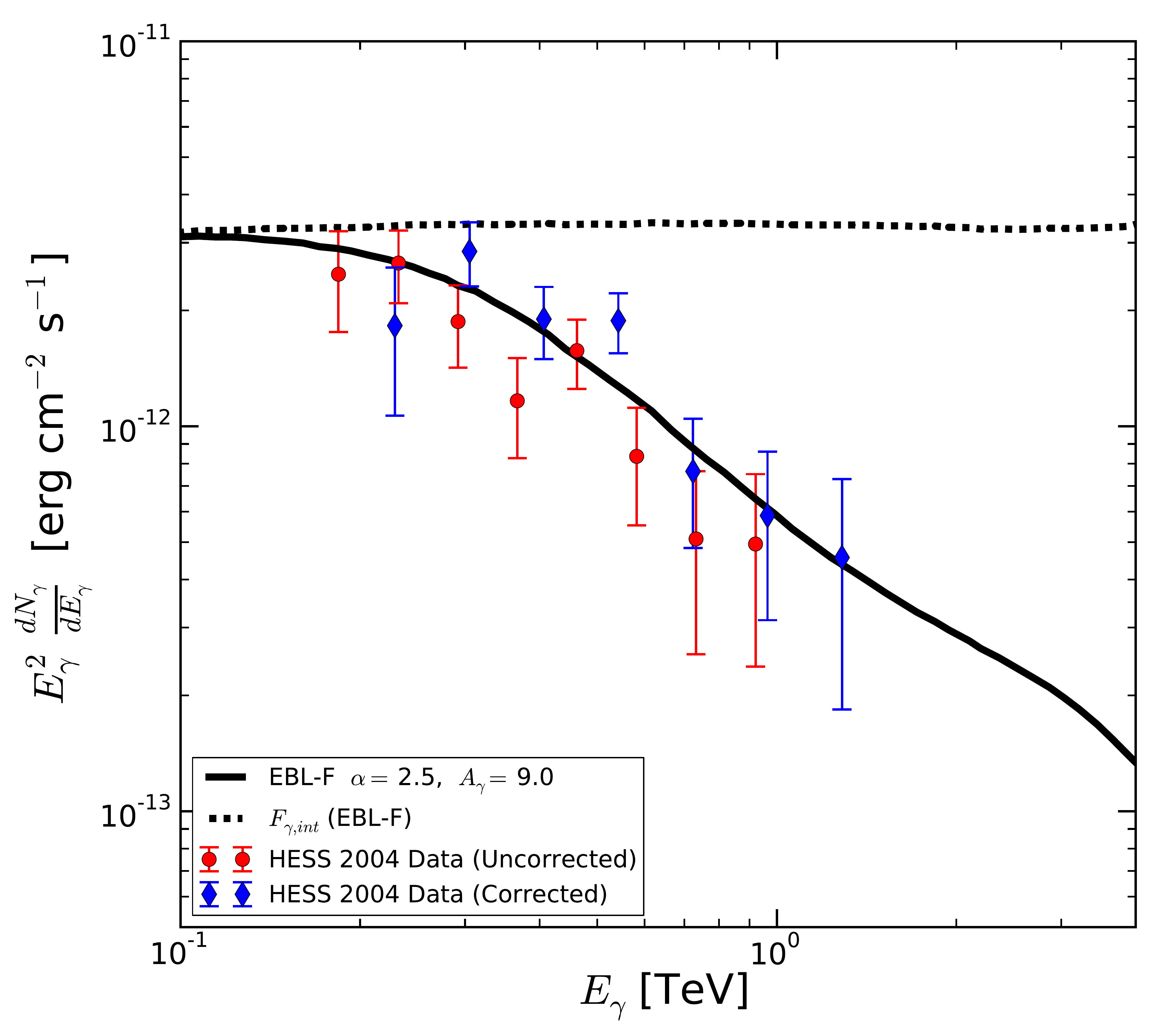}}
\par}
\caption{The VHE data of 2004 (uncorrected and corrected) from H
  2356-309 observed by HESS telescopes are shown for comparison.
Also the corrected data is fitted with the photohadronic model using
EBL-F deabsorption. The predicted intrinsic flux is also shown.
\label{fig:309OnlyFran04}
}
\end{figure}

\begin{figure}
{\centering
\resizebox*{0.5\textwidth}{0.35\textheight}
{\includegraphics{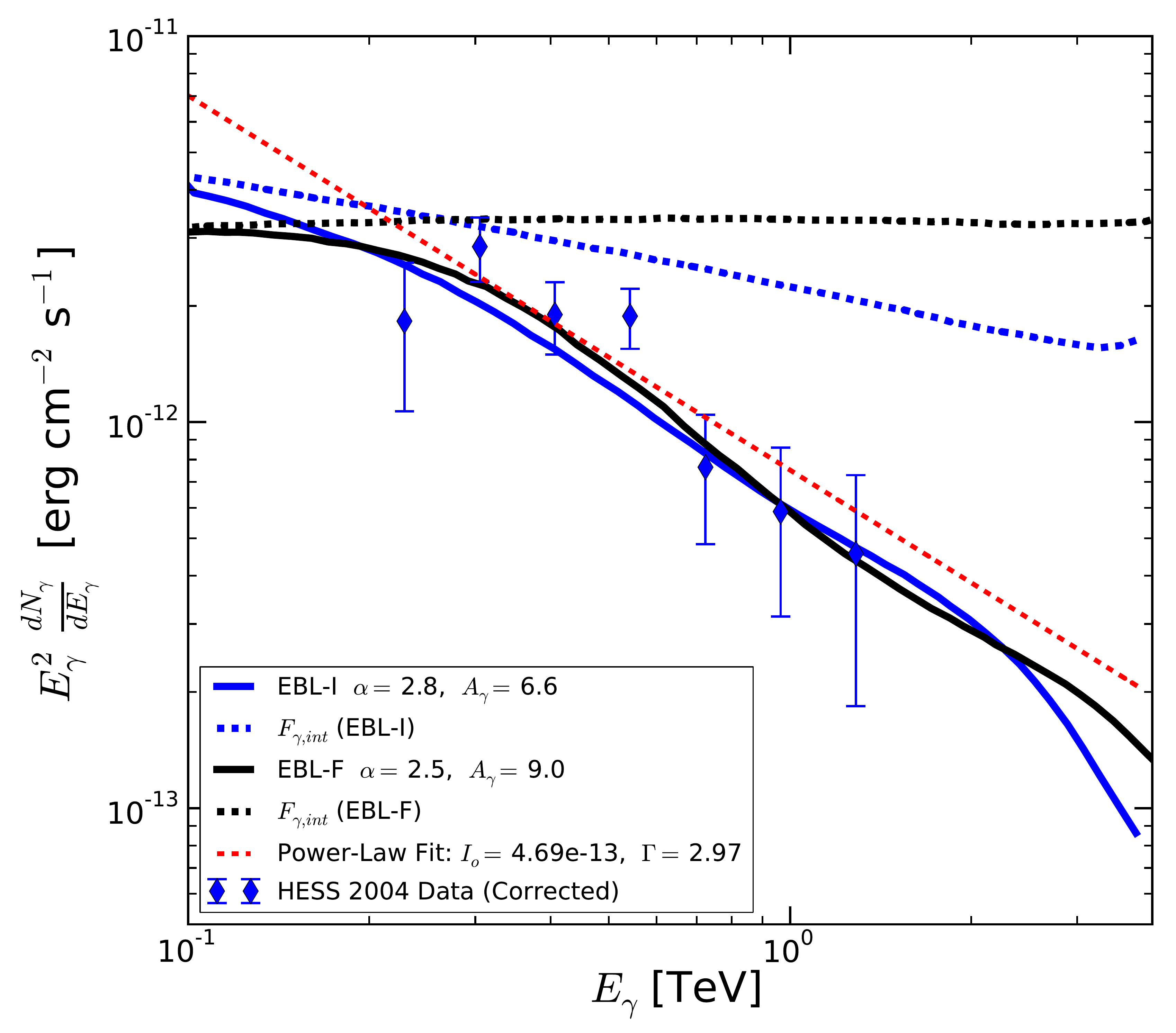}}
\par}
\caption{The fit to the corrected data of 2004 using
  different EBL models are shown. Also the power-law fit is shown for comparison.
\label{fig:309fit04C}
}
\end{figure}

\begin{figure}
{\centering
\resizebox*{0.5\textwidth}{0.35\textheight}
{\includegraphics{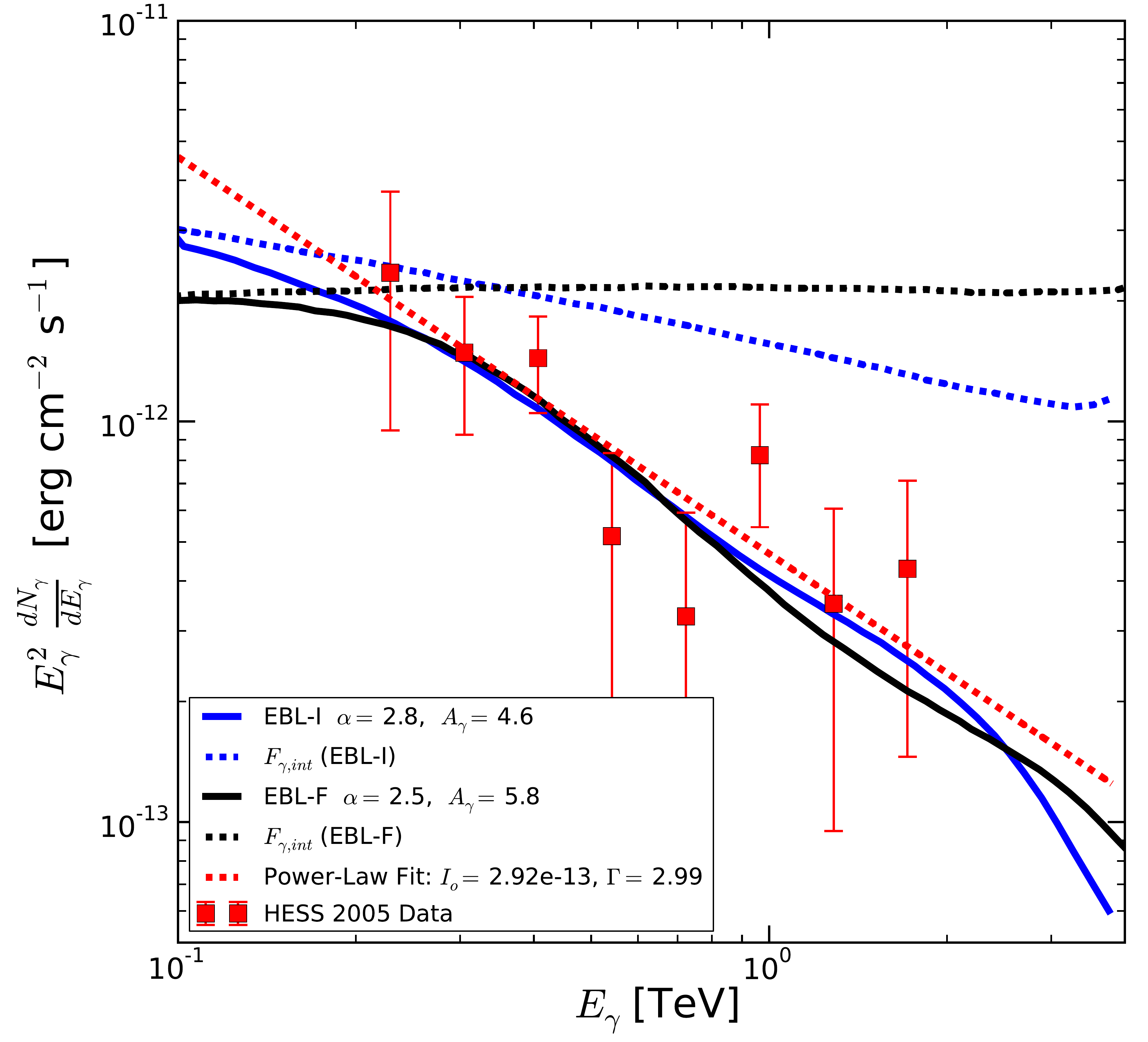}}
\par}
\caption{The VHE data of 2005 is fitted by different EBL models.
A power-law fit to the data is shown for comparison.
\label{fig:309fit05}
}
\end{figure}

\begin{figure}
{\centering
\resizebox*{0.5\textwidth}{0.35\textheight}
{\includegraphics{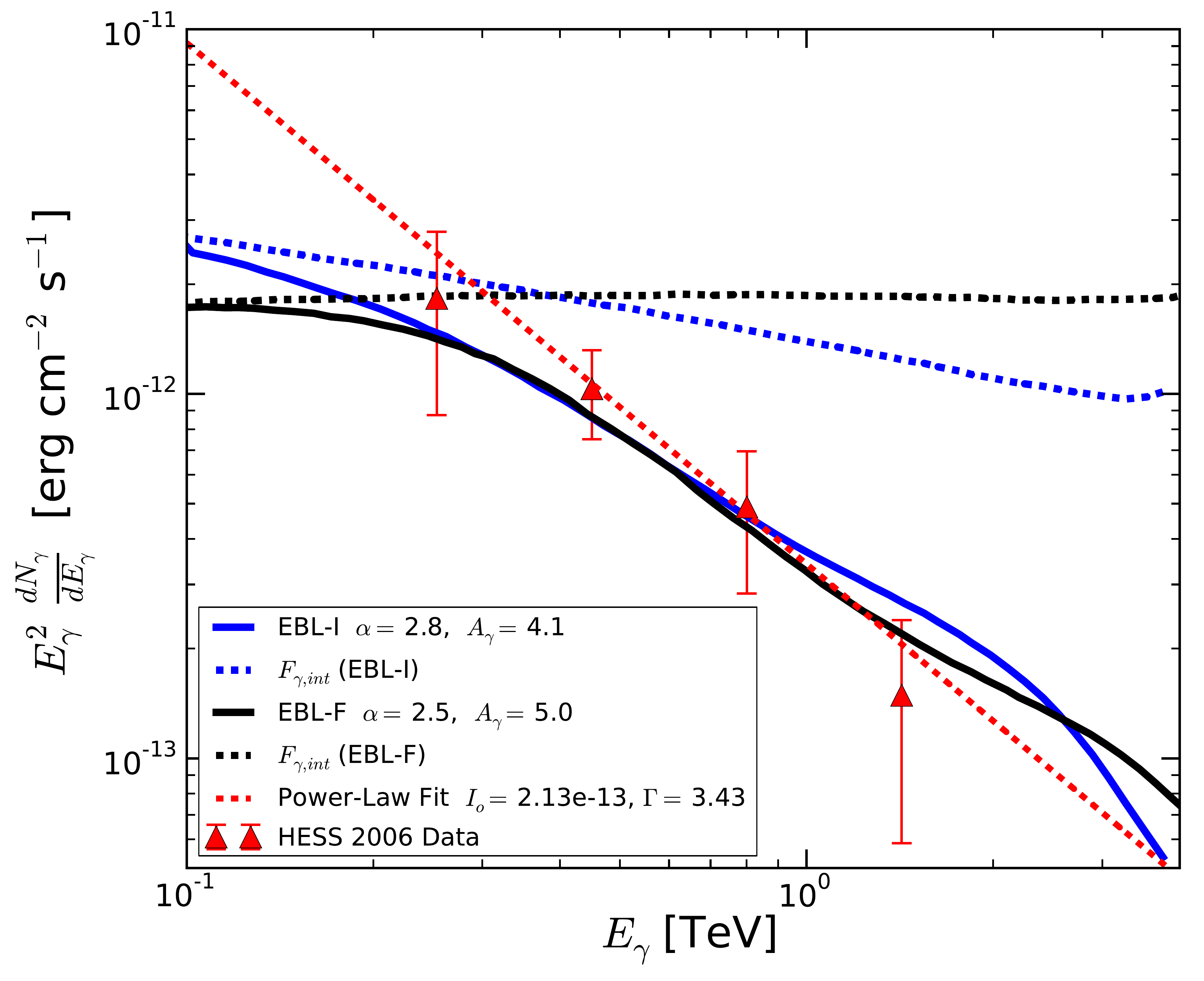}}
\par}
\caption{The VHE data of 2006 is fitted by different EBL models.
A power-law fit to the data is shown for comparison.
\label{fig:309fit06}
}
\end{figure}


\subsubsection{Correction to 2004 data}

The aging of the HESS detector and the accumulation of
dust on the optical elements of the telescopes affect the optical
efficiency of the detector system and it can reduce the efficiency by 
about 26\%  for the entire data sample. So the HESS collaboration 
reanalyzed the previously published result of 2004\cite{Aharonian:2005gh}
and added results of new observations from 2005 to 2007
in another article\cite{Abramowski:2010kg}. As a consequence of the
above correction the individual 
event energy is renormalized and correspondingly the flux changed.
The corrected 2004 integral flux is $\sim 50\%$ higher then the
original data. The observed VHE $\gamma$-rays of 2004 shifted from  
$0.18\, TeV \le E_{\gamma}\le 0.92\, TeV$ to $0.228\, TeV \le
E_{\gamma}\le 1.286\, TeV$. This new range of $E_{\gamma}$ shifted the
seed photon energy range to $5.94\, MeV \,(1.44\times 10^{21}\, Hz)\le
\epsilon_{\gamma}\le 33.5.5\, MeV\,(8.1\times 10^{21}\, Hz)$. We have shown both the uncorrected and
corrected data of 2004 for comparison in Fig. \ref{fig:309OnlyFran04}.
The corrected data is fitted well by photohadronic model with EBL-F
(black curve) correction for $\alpha=2.5$ and $A_{\gamma}=9.0$. By comparing these values
with the corresponding parameters of Fig. \ref{fig:309eblFran}, we
observe that the value of $A_{\gamma}$ has increased by $\sim\,27\%$
which implies an overall increase in the observed flux and the intrinsic flux by the
same amount with no other changes. 

We also fit the corrected data  with the EBL-I (blue curve)
correction to the photohadronic model and compare with other fits in
Fig. \ref{fig:309fit04C}. A good fit is obtained for $\alpha=2.8$ and
$A_{\gamma}=6.6$. Again, this new  $A_{\gamma}$
corresponds to a 10\% increase in the flux compared to the original
fit. 
A power-law fit (red dotted curve) with
$I=I_0 \,E^{-\Gamma}_{\gamma,TeV}$, where $\Gamma=2.97$ and $I_0=4.69\times 10^{-13}\, erg\,
cm^{-2}\,s^{-1}$ \cite{Abramowski:2010kg}, is shown for comparison. 
Although the EBL-F and EBL-I fits to the observe data are similar, for
$E_{\gamma} < 200$ GeV and $E_{\gamma} > 2$ TeV we can see a difference in their
behavior. Also both these fits are different from the power-law fit. 

We have also fitted the 2005 and 2006 data using the EBL-F (black
curve), EBL-I (blue curve) and
a power-law (red dotted curve) for comparison in Figs. \ref{fig:309fit05} and
\ref{fig:309fit06} respectively. Due to low photon statistics of 2007
observation, no spectrum was generated. In EBL-F good fits are 
obtained for same $\alpha=2.5$ but $A_{\gamma}=5.8$ for 2005 data and $A_{\gamma}=5.0$ 
for 2006 data respectively. Similarly for EBL-I good fits are obtained for same 
$\alpha=2.8$ but $A_{\gamma}=4.6$ for 2005 data and $A_{\gamma}=4.1$ 
for 2006 data respectively. The same value of $\alpha$ for a
particular model and data of different periods clearly shows that same 
acceleration mechanism is involved to accelerate the protons for the
observed VHE $\gamma$-rays from 2004 to 2006. The power-law fits to
2005 and 2006 data have similar behavior as the photohadronic fits in the
observed energy range.

There is no way to directly measure the photon density in the inner
compact region in the observed VHE ranges. 
Nonetheless, by assuming the scaling behavior of the photon densities for
different energies in the inner and the outer jets as shown in
Eq.(\ref{densityratio}), we relate the unknown densities of the inner
region with the known one in the outer region. In the outer jet
this range of $\epsilon_{\gamma}$ lies in the low energy tail region
of the SSC band and the sensitivity of the
currently operating $\gamma$-ray detectors are  not good enough to
detect these photons. 

The hidden jet has a size $R'_f < R'_b=7.5\times 10^{15}$ cm and here 
we take $R'_f\sim 10^{15}$ cm. Also 
by assuming the central black hole has a mass of 
$M_{BH}\sim 10^9\, M_{\odot}$ and using
the constraint on the highest energy
proton flux and the maximum luminosity of the inner jet to be smaller
than the Eddington luminosity, the $p\gamma$ optical depth satisfies
$0.005 \ll \tau_{p\gamma} \ll 0.097$. For our estimate we take
$\tau_{p\gamma}=0.01$ which gives the photon density in the inner jet
region $n'_{\gamma,f}\simeq 2\times 10^{10}\, cm^{-3}$.

\begin{figure}
{\centering
\resizebox*{0.5\textwidth}{0.35\textheight}
{\includegraphics{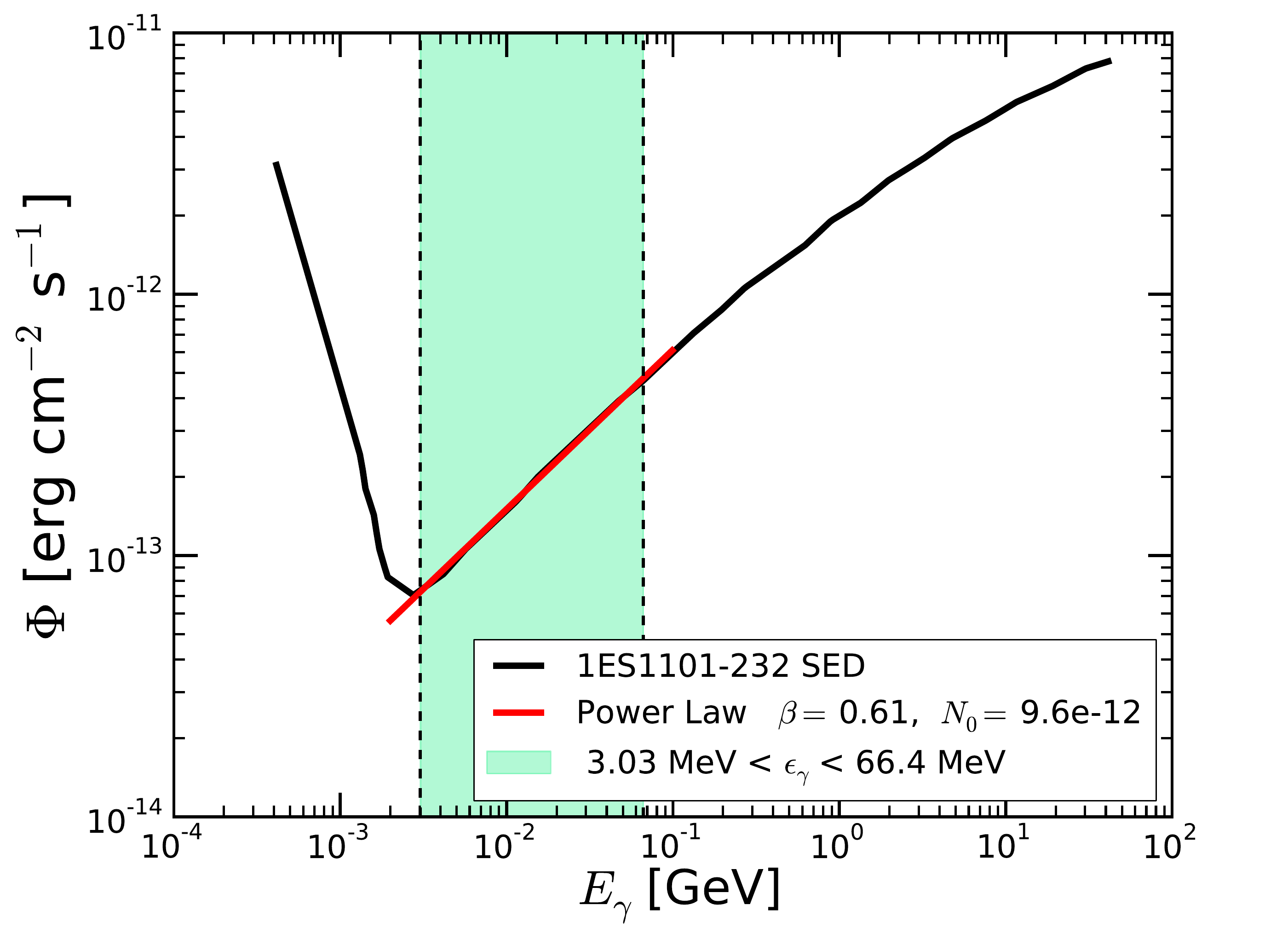}}
\par}
\caption{ 
The leptonic SED of the HBL 1ES 1101-232 is fitted with one-zone leptonic
model in ref. \cite{Costamante:2006pm} and 
the shaded region in the figure is the 
SSC energy range $3.03 \, MeV (7.3\times 10^{20}\, Hz) \le
\epsilon_{\gamma} \le 66.4\, MeV (1.6\times 10^{22}\, Hz)$
corresponding to the VHE
$\gamma$-ray energy in the range $0.18\, TeV \le E_{\gamma}\le 2.92\,
TeV$ of the HBL 1ES 1101-232. The corresponding SSC flux $\Phi_{SSC}$ in y-axis
is fitted with a power-law
$\Phi_{SSC}=N_0 E^{-\beta}_{\gamma,{\text {TeV}}}$, where $N_0=9.6\times 10^{-12}\,
erg\, cm^{-2}\, s^{-1}$ and $\beta=0.61$. 
\label{fig:232sscpowerlaw}
}
\end{figure}

\begin{figure}
{\centering
\resizebox*{0.5\textwidth}{0.35\textheight}
{\includegraphics{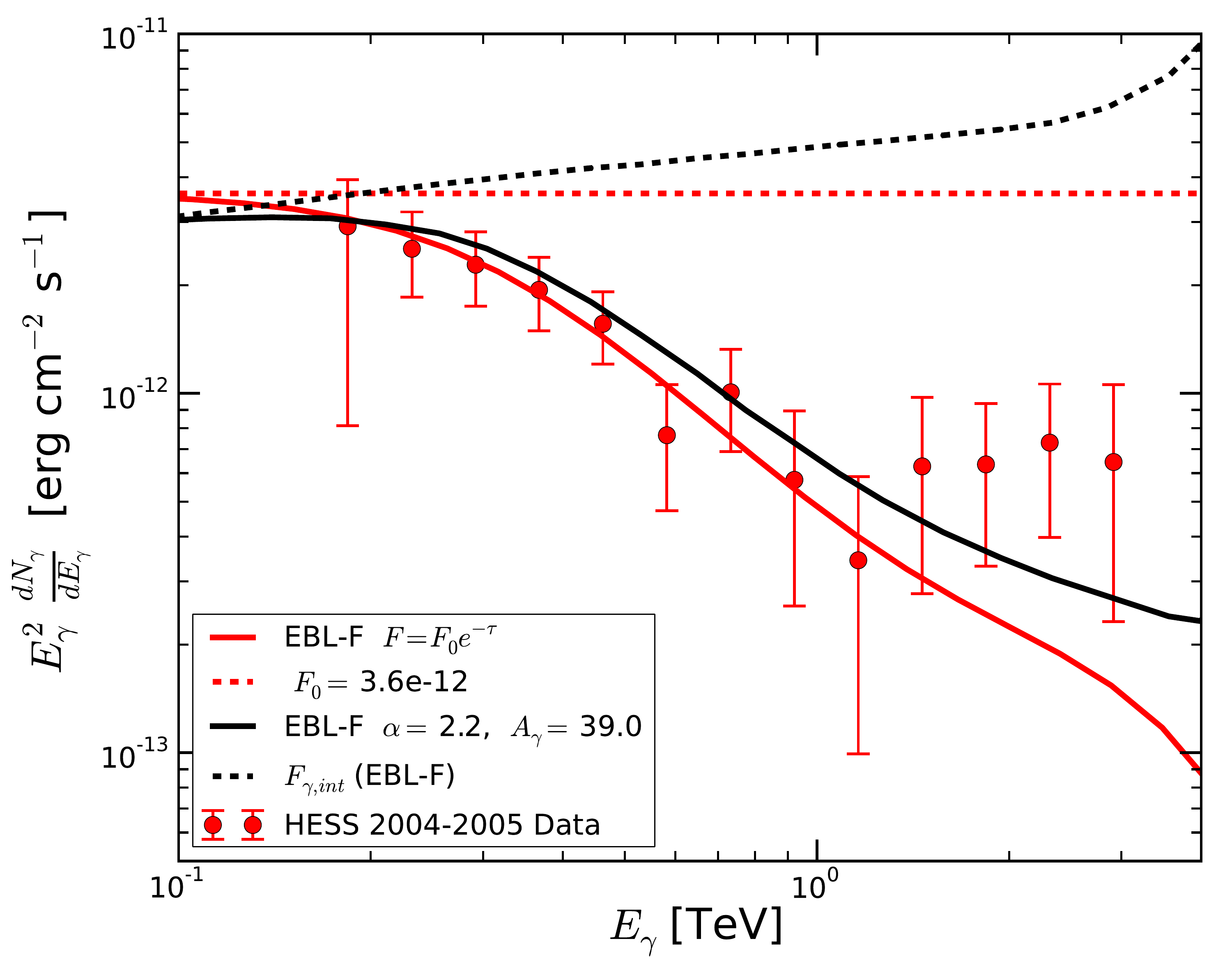}}
\par}
\caption{The VHE emission from HBL 1ES 1101-232 in 2004 and 2005 observation
  by HESS telescopes is shown along with the rescaling of the
  EBL-F attenuation factor and EBL-F correction to the photohadronic
  model. The intrinsic fluxes are also shown. 
\label{fig:232eblFran}
}
\end{figure}

\begin{figure}
{\centering
\resizebox*{0.5\textwidth}{0.35\textheight}
{\includegraphics{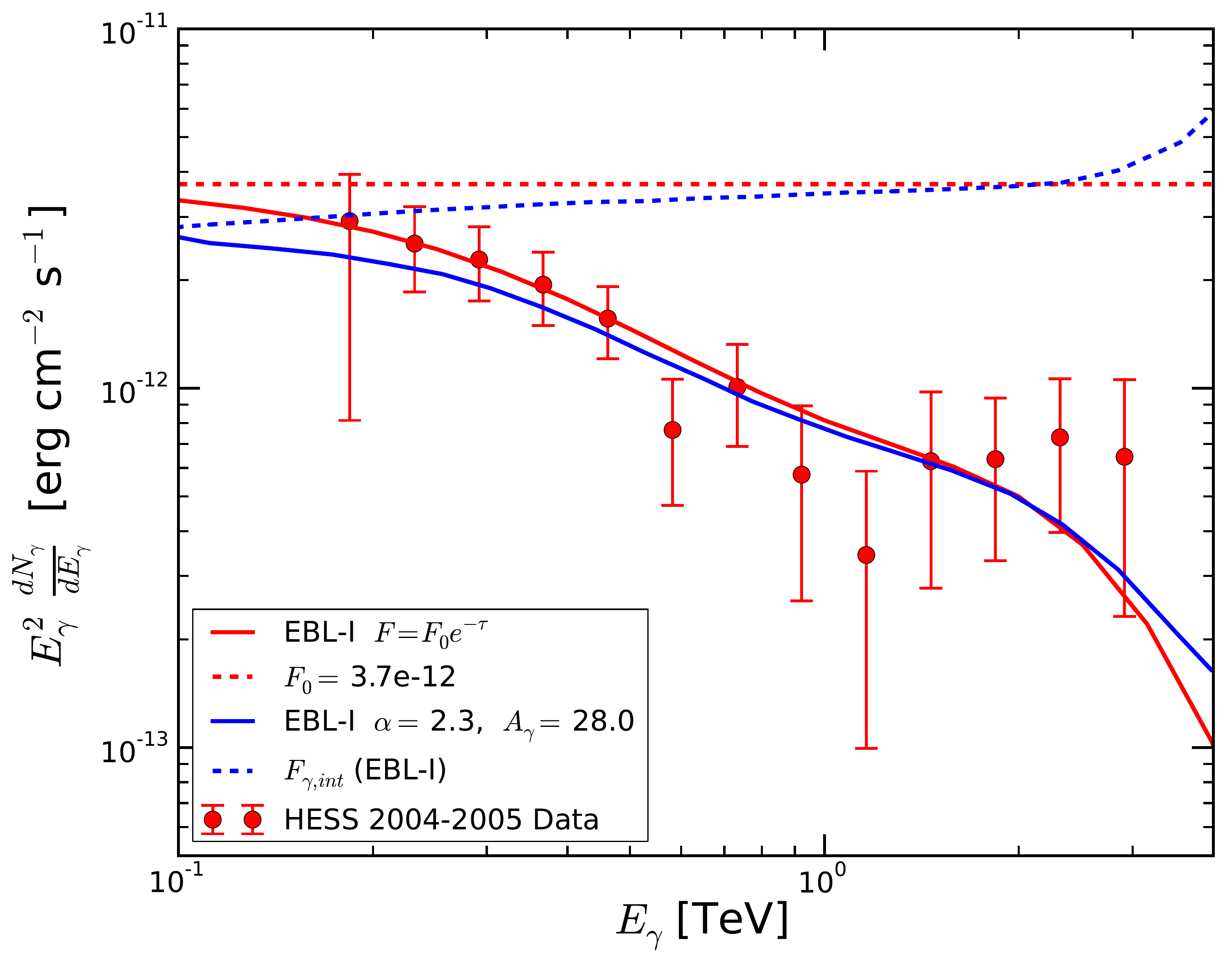}}
\par}
\caption{The VHE emission from HBL 1ES 1101-232 in 2004 and 2005 observation
  by HESS telescopes is shown along with the rescaling of the
  EBL-I attenuation factor and EBL-I correction to the photohadronic
  model. The intrinsic fluxes are also shown. 
\label{fig:232eblInoue}
}
\end{figure}



\begin{figure}
{\centering
\resizebox*{0.5\textwidth}{0.35\textheight}
{\includegraphics{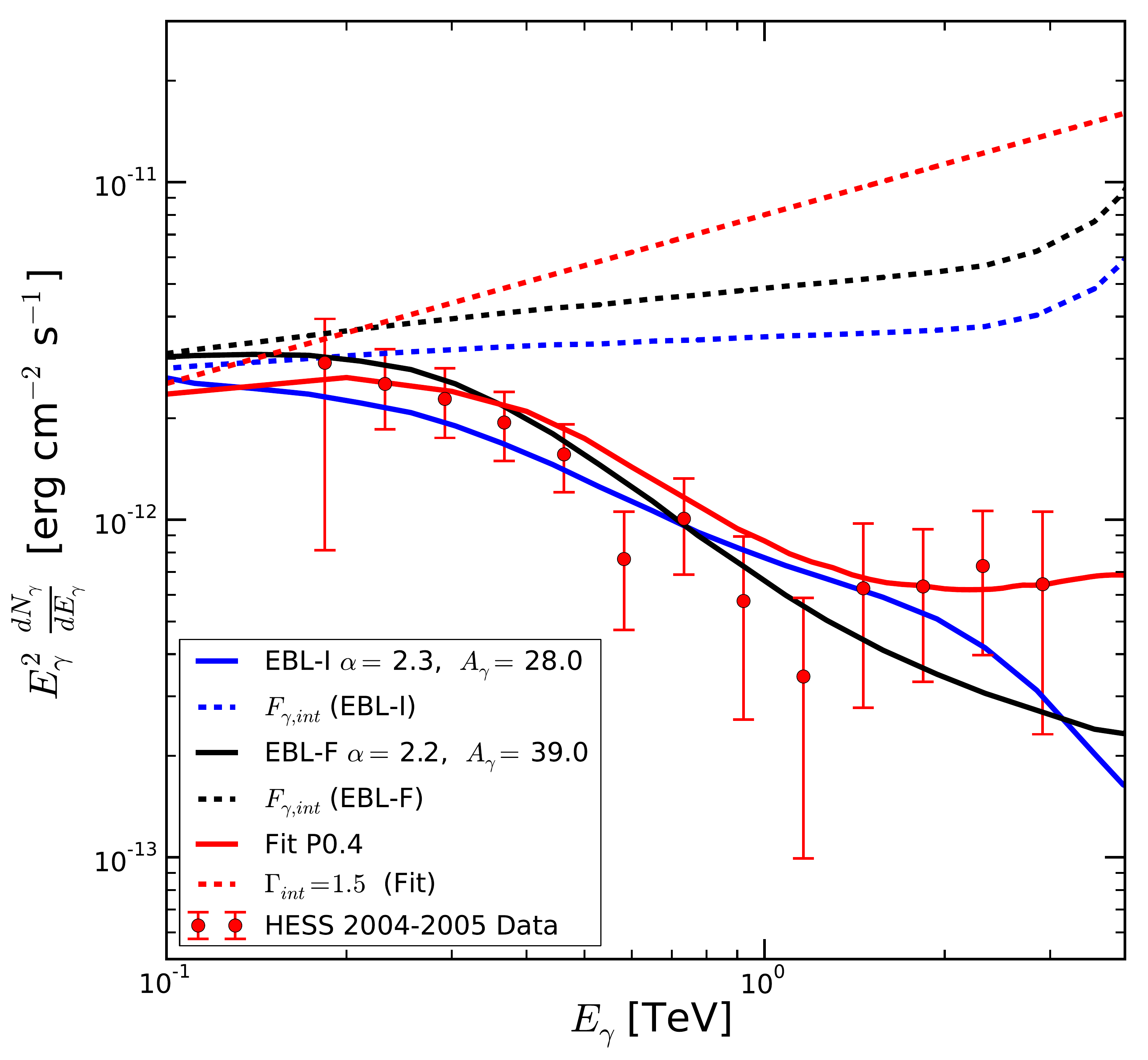}}
\par}
\caption{The observed VHE flux from 1ES 1101-232 is fitted using EBL-F and EBL-I to the
  photohadronic model. Also Aharonian et al. fit to the data using
  P0.4 is shown for comparison. The intrinsic flux predicted by all
  these models are also shown.
\label{fig:SED232}
}
\end{figure}


\subsection{1ES 1101-232}
The HBL 1ES 1101-232 resides in an elliptical
host galaxy at a redshift of z = 0.186\cite{Remillard:1989,Falomo:1994a}. 
The radio maps of this HBL show an one-sided, not
well-collimated jet structure at a few kpc distance from the
core\cite{Laurent:1993}. In 2004 and 2005, 1ES 1101-232 was
observed by the HESS telescopes and following the detection of a weak
signal in its observations, an extended multifrequency
campaign was organized for 11 nights in March 2005 to study the
multiwavelength  emission and to look for possible
correlated variability in different wavebands\cite{Aharonian:2005gh}. The exposure time
for VHE observation was approximately 43 hours. Also simultaneous
observations were carried out in X-rays by RXTE, and in optical with
ROTSE 3c robotic telescope. However, there were no simultaneous
observation in GeV energy range. But, the analysis of 3.5 years data
collected from August 2008 to February 2012 by Fermi-LAT reported 
the observation of GeV emission from this object
\cite{Finke:2013tyq}. From the simultaneous observations in
optical, X-rays and VHE $\gamma$-rays,
Aharonian et al. constructed a truly simultaneous SED of 1ES
1101-232. In 2006 May and July, Suzaku observed this HBL in
X-rays which was also quasi-simultaneously observed with HESS and MAGIC
telescopes and no significant variability was observed neither in
X-rays nor in $\gamma$-rays\cite{Aharonian:2007nq}. In fact during this observation period it
was found in a quiescent state
with the lowest X-ray flux ever measured. The multiwavelength
observation of the blazar 1ES 1101-232 during the flaring in 2004-2005 is
used to constructed the synchrotron and SSC SED using one zone
leptonic model\cite{Aharonian:2007nq,Costamante:2006pm} 
and the parameters for the best fit are given in Table-\ref{tab:tab1}. For our
calculation we shall use these parameters.

The observed VHE flare of 1ES 1101-232 was in the energy range $0.18\,
TeV \le E_{\gamma} \le 2.92\, TeV$. This corresponds to seed
photon energy range $3.03 \, MeV (7.3\times 10^{20}\, Hz) \le
\epsilon_{\gamma} \le 66.4\, MeV (1.6\times 10^{22}\, Hz)$ in the
inner jet region, where the Fermi accelerated protons
in the energy range $1.8\, TeV \le E_p \le 30 \, TeV$ collide to
produce $\gamma$-rays and neutrinos  through intermediate
$\Delta$-resonance and pions.  This range of $\epsilon_{\gamma}$
corresponds to the low energy tail of the SSC band. In the energy
range $3.03 \, MeV \le \epsilon_{\gamma} \le 66.4\, MeV$ we observed
that the SSC flux can be fitted with a power-law (red line) given as
$\Phi_{SSC}=9.6\times 10^{-12}\,erg\,
cm^{-2}\,s^{-1} E^{-0.61}_{\gamma,TeV}$ which is shown in Fig. \ref{fig:232sscpowerlaw}. 

In Fig. \ref{fig:232eblFran},  we rescale the attenuation factor of
EBL-F by $F_{\gamma,int}=3.6\times 10^{-12}\, erg\,
cm^{-2}\, s^{-1}$ to fit the observed VHE data (red curve). It shows that
the rescaling can't fit the VHE data above 1.5 TeV. However, 
a good fit to the VHE flare data is obtained for
$\alpha=2.2$ and $A_{\gamma}=39.0$ in the photohadronic model with
EBL-F (black curve )
correction and this corresponds to an intrinsic spectrum with
$\alpha_{int}=1.81$.

Again by multiplying $F_{\gamma,int}=3.7\times 10^{-12}\, erg\,cm^{-2}\,
s^{-1}$ to the attenuation factor of EBL-I we can fit well the observed
data below 1 TeV. However, above 1 TeV the fitted curve differs from
the observed data as shown in Fig. \ref{fig:232eblInoue} (red curve). In the same figure we have
also shown the photohadronic model with the EBL-I correction fit (blue
curve) to the observed data for $\alpha=2.3$ and $A_{\gamma}=28.0$.
The photohadronic fit almost coincides with the rescaling of the
attenuation factor and having $\alpha_{int}=1.91$ which is softer than
the one by EBL-F.

In Fig. \ref{fig:SED232}, we have compared all these models and the fit
of ref.\cite{Aharonian:2005gh}. Rescaling the originally normalized EBL by 40\%
(P0.4, red curve), Aharonian et al. could fit the data well which is shown in
the figure. At the same time the photohadronic model accompanied by
EBL-F (black curve) and EBL-I (blue curve) also fit the data well. But all these three fits
behave differently in the high energy regime. While the fit in
ref.\cite{Aharonian:2005gh} 
slightly increases beyond $\sim 2$ TeV, the EBL-I predict a
drop in the flux above this energy limit and EBL-F flux is relatively
shallow. Even though the overall fit to the observed data by different
models are similar, their corresponding intrinsic fluxes behave differently. 
The fit by Aharonian et al. gives the intrinsic
spectral index $\alpha_{int}=1.5$ which is hard (red dotted curve). However, in the
photohadronic model with EBL-F we have $\alpha_{int}=1.81$
and with EBL-I it gives $\alpha_{int}=1.91$. So the photohadronic
scenario gives milder intrinsic spectral index compared to the
one by ref. \cite{Aharonian:2005gh} in their original fit. 

In 1ES 1101-232, the BH process will produce leptons with energies below
$<\, 30$ TeV and  synchrotron emission from these electrons and
positrons in $0.1$ G magnetic field in the inner jet region will produce
synchrotron photons below 3 MeV energy range. Thus, these photons
will not contribute for the enhancement of the photon flux in the low
energy tail region of the SSC band.

We have also calculated the photon density in the inner jet
region. For this we have taken the central black hole mass
$M_{BH}\sim 10^9\,M_{\odot}$ and the inner jet region has a size
$R'_f\sim 10^{15}$ cm. Using the constraint on the highest energy
proton flux and the maximum luminosity of the inner jet to be smaller
than the Eddington luminosity we get
$0.001\ll \tau_{p\gamma} \ll 0.29$. We take $\tau_{p\gamma}\sim 0.01$
which gives $n'_{\gamma, f} \simeq 2\times 10^{10}\, cm^{-3}$ in the inner
jet region.

\section{Summary}

The multi-TeV emission from the HBLs H 2356-309 and 1ES 1101-232 were
observed by HESS telescopes during 2004 to
2007. For the first time, the VHE observation from
H 2356-309 in 2004 and in 2004-2005 from 1ES 1101-232 were analyzed by
Aharonian et al.\cite{Aharonian:2005gh} to derive strong upper limits on the EBL
which was found to be consistent with the lower limits from the integrated
light of resolved galaxies. While the intrinsic spectrum of H
 2356-309 found to be flat,  for 1ES 1101-232 it was hard
 $\alpha_{int} \leq 1.5$.  
Here we have used the photohadronic
model accompanied by two template EBL
models EBL-F and EBL-I to fit the observed VHE data from these two HBLs and to predict
their intrinsic spectra. Although the blazar jet environment plays an
important role in attenuating the VHE $\gamma$-rays, the absorption of it
within the jet is neglected by assuming that the intrinsic flux takes
care of this extraneous effect. 

An important ingredient for the
photohadronic scenario is the SSC flux $\Phi_{SSC}$.
From the simultaneous multi-wavelength observations of these HBLs,
one-zone leptonic models are constructed to fit the observed
data well and the resulting
parameters and $\Phi_{SSC}$ are used here for the analysis of our
results. In the photohadronic model the intrinsic flux $F_{\gamma,int}\propto
E^{-(\alpha+\beta-3)}_{\gamma}$ and the power index $\beta$  is
fixed for a given leptonic model. So the proton spectral index $\alpha$
is the only free parameter here.

A good fit for  the 2004 corrected VHE spectrum of H 2356-309 is
achieved by photohadronic model with the EBL correction from EBL-F and
EBL-I. However, the intrinsic spectrum is different for each EBL
model. While the EBL-F correction gives a flat intrinsic spectrum,
a softer intrinsic spectrum is obtained with the EBL-I correction.
The same spectral index $\alpha$ of the respective EBL
model but different normalization can fit the VHE spectra of 2005 and
2006 well.

The multi-TeV spectrum of 1ES 1101-232 is also fitted using the EBL-F
and EBL-I and compared with the original fit by Aharonian et
al. The overall fit to the observed VHE SED by all these models are
similar but their corresponding intrinsic spectra are different. The
$\alpha_{int}$ of the EBL-I is softer then the  EBL-F which is again
softer than the fit by ref. \cite{Aharonian:2005gh}. 
In future, for a better understanding of the EBL
effect and the role played by the SSC photons on the VHE
$\gamma$-ray flux from intermediate to high redshift blazars, it is necessary
to have simultaneous observations in multi-wavelength to the flaring
objects and to accurately model the low energy SSC tail region. 

We thank D. Khangulyan, Yoshiyuki Inoue,
Susumu Inoue, Vladimir L. Y\'{a}\~{n}ez, M. V. Barkov and Haoning He for many useful
discussions. S.S. is a Japan Society for the Promotion of Science
(JSPS) invitational fellow. The work of S. S. is partially supported by
DGAPA-UNAM (Mexico) Project No. IN110815 and PASPA-DGAPA, UNAM. This work is partially
supported by RIKEN, iTHEMS \& iTHES Program and also by Mitsubishi Foundation.


\end{document}